\documentclass[aps,prr,twocolumn,groupedaddress]{revtex4-2}

\usepackage{amsmath}
\usepackage{amssymb}
\usepackage{accents}
\usepackage{contour}
\usepackage{relsize}
\usepackage{mathtools}
\usepackage{dsfont}
\usepackage{natbib}
\usepackage[hidelinks]{hyperref}
\usepackage{tikz}
\usepackage[bb=stix,bfbb]{mathalpha}
\usepackage{bm}
\usepackage{graphicx}
\usepackage{placeins}
\usepackage{subcaption}
\usepackage{accents}

\graphicspath{{imgs/}}

\newcommand*{\dt}[1]{%
	\accentset{\mbox{\large\bfseries .}}{#1}}

\newcommand{\vfield}[1]{
	\boldsymbol{#1}
}
\newcommand{\pfield}[1]{
	\mathrm{#1}
}
\newcommand{\inv}[1]{
	\mathrm{#1^{-1}}
}

\def\vecU{\boldsymbol{U}}
\def\vecUb{\vfield{U^{0}}}
\def\vecPb{\pfield{P^{0}}}
\def\vecu{\vfield{u}}
\def\vecv{\vfield{v}}
\def\vecw{\vfield{w}}
\def\vecy{\vfield{y}}
\def\vecubar{\bar{\vecu}}
\def\vecvbar{\bar{\vecv}}
\def\vecwbar{\bar{\vecw}}
\def\vecybar{\bar{\vecy}}
\def\bcdot{\boldsymbol{\cdot}}
\def\ie{\textit{i.e.} }

\def\etc{{etc.} }
\def\Lu{\mathbf{L}}
\def\Nu{\mathbf{N}}
\def\invRec{\inv{\Rey_{c}}}
\def\Rey{\mathrm{Re}}

\def\parv{z}
\def\vecparv{\mathbf{\parv}}

\begin{document}

\title{A model for limit-cycle switching in open cavity flow}
\author{Prabal S. Negi}
\email[]{prabal.negi@oist.jp}
\affiliation{Okinawa Institute of Science and Technology Graduate University, Onna, Okinawa 904-0495, Japan}

\date{\today}

\begin{abstract}
	A reduced mathematical model for the flow in an open cavity is presented. The reduction is based on the center manifold theory applied to a perturbation of the original system which allows for a codimension two bifurcation point. The model exhibits many of the key characteristics observed in the flow dynamics including unstable quasi-periodic edge states as well as switching of limit cycles with parameter variations. An explanation for the exchange of stabilities of the limit-cycles is presented based on the cross-coupling terms of the two amplitude equations.
\end{abstract}

\maketitle

\section{Introduction}
\label{sec:intro}

The two-dimensional shear driven flow over a cavity presents an interesting case of successive bifurcations appearing in a hydrodynamic flow. The flow case has been brought to the attention of the hydrodynamic stability community by \cite{sipp07}, where, the authors used the geometry to investigate the theoretical aspects of stability analysis around time averaged mean flows. It has featured as an object of investigation in various different contexts of stability and control \citep{rowley06,sipp10,barbagallo09}, model reduction \citep{loiseau18}, self-consistent modeling \citep{meliga17}, center manifold reduction \citep{negi24} \etc

The basic scenario of the case is this - a boundary layer flow is allowed to develop over flat plate which contains a large depression in the form of a square cavity. Up to a Reynolds number of roughly $\Rey\approx4130$ (based on the cavity height and freestream velocity) a steady circulation develops within the cavity and the developing boundary layer flows smoothly over this circulation as it crosses the open cavity. The first bifurcation of the flow then occurs and the flow settles on to a low amplitude limit-cycle oscillation (LCO) with a characteristic frequency and spatial wavelength. Subsequently, at around $\Rey\approx4500$ a second bifurcation seems to occur and a distinctly different oscillation frequency and wavelength becomes dominant in the flow. While classic asymptotic methods are able to capture the characteristics of the first bifurcation \citep{sipp07,negi24}, modeling the second bifurcation has been a challenge. The second limit-cycle also appears to behave in a manner consistent with Landau's theory of saturation provided by the mean flow correction \citep{landau_52}, albeit one that requires second order terms to be taken into account \citep{meliga17}. However no attempt has been made so far to model both the limit cycles together. 

A rather interesting and exhaustive investigation of the flow dyanamics within this Reynolds number regime has been performed by \cite{bengana19}. The authors employed several tools within the dynamical systems framework - linear stability, Floquet analysis, mean flow stability analysis and edge tracking to build a comprehensive picture of the successive bifurcations in the flow as the Reynolds number is varied. Besides the two distinct limit cycles, the authors were also able to identify a quasi-periodic state which has been interpreted as a non-linear superposition of the two distinct limit cycles. This quasi-periodic state is the edge state between the two limit cycles although, the authors speculate that the state might in fact be periodic with a very long period. Various bifurcation points where qulitative changes in flow dynamics are expected to occur are also identified.  

Despite the detailed analyses of previous studies, a reduced model representing the essential dynamics of the problem has remained out of reach. In \cite{bengana19} the authors propose a normal form representation of the dynamics but do not attempt to derive the representation or specify the coefficients. The current work proposes a reduced representation based on the center manifold theory \citep{carr82,carr83b,wiggins03,guckenheimer83,roberts14}. At the bifurcation point one could evaluate the center subspace of the linearized operator, and the center manifold as the (nonlinear) continuation of this tangent subspace. However, this system exhibits only a single oscillatory mode in the center subspace. This is obviously insufficient for the representation of the dynamics where two distinct limit-cycle oscillations can emerge. Instead, we consider a perturbed system with a codimension two bifurcation point via the introduction of a ``pseudo-parameter''. This is fairly straightforward to construct whenever the relevant direct and adjoint tangent vectors are known. The center manifold can now be obtained for the perturbed system asymptotically, with the asymptotic variables also including the new pseudo-parameter in addition to the modal variables (and inverse Reynolds number). The original system is then approximated by replacing the pseudo-parameter by the appropriate value. The reduced model is analyzed and its predictions compared to the extensive analysis reported in \cite{bengana19}. The approach can be thought of as an example of ``backward theory'' developed by \cite{hochs19,roberts22}, wherein, the dynamics of the original system on an invariant manifold are approximated by a nearby system's invariant manifold. Here though, we do not construct the exact invariant manifold but rather an asymptotic one.


\section{Open Cavity Flow}
\label{sec:cavity_setup}

\begin{figure}
	\centering
	\includegraphics[width=0.49\textwidth]{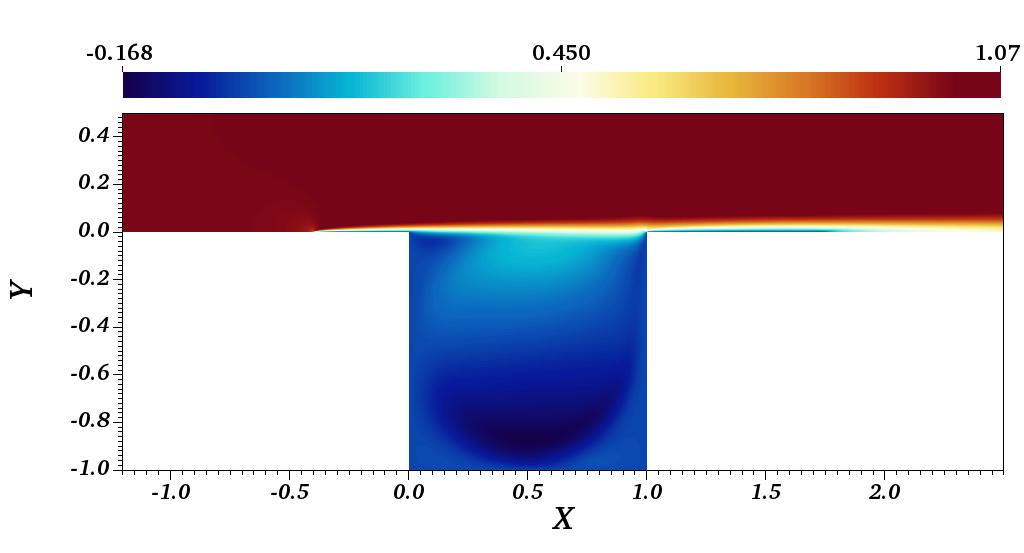}
	\includegraphics[width=0.40\textwidth]{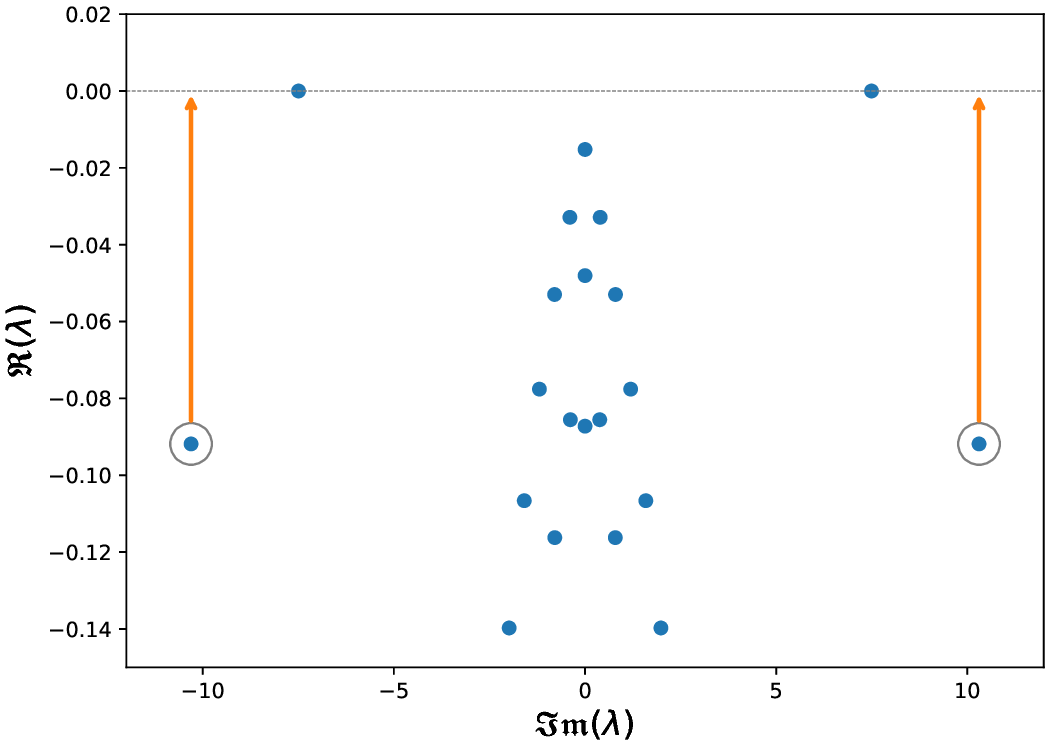}
	\caption{(Top) streamwise velocity of the stationary base flow at $\Rey_{c}=4131.33$ and (bottom) the spectrum (blue dots) obtained at the bifurcation point.}
	\label{fig:cavity_base_spectrum}
\end{figure}

The setup of the two-dimensional open cavity flow has been described in \cite{negi24}, which follows from that of \cite{sipp07}. Briefly, the domain consists of a square cavity with sides of length one and an open channel flow is constructed above the cavity. The open channel has a width equal to half the cavity length, and a symmetry boundary condition is applied to the upper boundary of the channel. A uniform (streamwise) velocity boundary condition of $u=1$, is applied to the inlet located at $x=-1.2$. A symmetry boundary condition is applied to the lower wall of the channel from the inlet to $x = -0.4$ which allows the flow to develop freely from the inlet. A no-slip condition is applied on the lower wall thereafter uptill  $x = 1.75$ (including the cavity walls), after which, a symmetry condition is applied again from $x = 1.75$ to the outlet, located at $x = 2.50$. All physical quantities are normalized using the inlet velocity, fluid density and the cavity length. The continuous system is descretized using the spectral-element-method \citep{patera84} and the Nek5000 code \citep{nek5000} is utilized for computations. The overall geometry can be seen in figure~\ref{fig:cavity_base_spectrum} (top). 

The critical point for the particular geometry is found to be $Re_{c}=4131.33$. There is a slight variability in the literature on the value of the bifurcation point \citep{sipp07,meliga17,bengana19}. However all reported values are within $1\%$ error of each other, and the calculated value here falls within that range. The base flow at bifurcation is shown in figure~\ref{fig:cavity_base_spectrum} (top). The calculated baseflow velocities will be denoted $\vecUb = [\vecUb_{x}; \vecUb_{y}]$ (which includes its two components) and, its associated pressure field is given by, $\vecPb$. Hereafter we will only deal with the deviation from this base flow velocities $\vecu = [\vecu_{x};\vecu_{y}]$ and pressure $\pfield{p}$, with the total velocity field being $\vecU = \vecUb + \vecu$, and the total pressure given by $\pfield{P} = \vecPb + \pfield{p}$. The governing equations for the velocity and pressure deviations are given by the Navier--Stokes. Considering the velocity and pressure together as one vector $\vecubar = [\vecu; \pfield{p}]$, we write this compactly as,
\begin{align}
	\label{eqn:compact_evolution1}
	\partial \vecubar/\partial t =& \Lu\vecubar + \Nu(\vecubar), \\
	\Lu =& \begin{bmatrix}
		\inv{\Rey}\nabla^{2} - (\vecUb\bcdot\nabla) -(\nabla\vecUb)\bcdot &  -\nabla \\
		\nabla\bcdot&  0
	\end{bmatrix}, \nonumber \\
	\Nu(\vecubar) =& \begin{bmatrix}
	- \vecu\bcdot\nabla\vecu \\
	0
	\end{bmatrix}, \nonumber
\end{align}
where, $\Lu$ and $\Nu$ are respectively the linear and non-linear operators at bifurcation. The calculated spectrum at bifurcation is shown in figure~\ref{fig:cavity_base_spectrum} (bottom), which was evaluated using the Krylov-Schur algorithm \citep{stewart02}. The spectral problem was also solved for the adjoint operator in order to obtain the adjoint eigenmodes. 

A single pair of complex-conjugate neutral modes can be observed in the figure~\ref{fig:cavity_base_spectrum}, lying on the $x-$axis. This pair of modes (numbered $1$ and $2$), with the eigenvalue $\lambda_{1,2} = \pm7.495\iota$, governs the dynamics of the first limit cycle that emerges as the system moves just past the bifurcation point \citep{negi24,bengana19}. However, deep in the stable spectrum lies another pair of modes with the eigenvalues $\lambda_{3,4}=-0.092\pm10.31\iota$, marked by the gray circles. This mode has been found to be responsible for the second limit cycle oscillation that becomes dominant with further increases in Reynolds numbers \citep{bengana19}. One can not directly incorporate this mode into the center manifold evaluation of the system given by equation~\ref{eqn:compact_evolution1}. However, if one could perturb the system so that modes $3$ and $4$ are also neutral simultaneously with modes $1$ and $2$, then one has a codimension two bifurcation point and the resulting center manifold would be the nonlinear continuation of the tangent space spanned by these two pair of modes. Such a system is easily built. Denoting the direct and adjoint eigenvectors for the modes as $\vecvbar_{i}$ and $\vecwbar_{i}$, respectively, with $i\in1,2,3,4$, one may build a synthetic system as,
\begin{align}
	\label{eqn:compact_evolution2}
	\partial \vecubar/\partial t = \widetilde{\Lu}\vecubar + \Nu(\vecubar), && \widetilde{\Lu} = \Lu + \sigma(\vecvbar_{3}\vecwbar^{\dagger}_{3} + \vecvbar_{4}\vecwbar^{\dagger}_{4})
\end{align}
where $^{\dagger}$ represents the complex-conjugated transpose and $\sigma$ is the new pseudo-parameter that has been introduced. The eigenvectors of the new linearized system $\widetilde{\Lu}$ are identical to those of $\Lu$. All eigenvalues except for $\lambda_{3,4}$ also remain unchanged. The new affected eigenvalues are $\widetilde{\lambda}_{3,4} = \lambda_{3,4} + \sigma$. Clearly, if one sets $\sigma = -\mathfrak{R}(\lambda_{3,4})$, \ie the real part of $\lambda_{3,4}$, then the eigenvalues of these modes gets mapped to the $x-$axis, as indicated by the arrows in figure~\ref{fig:cavity_base_spectrum}, and we obtain a system with a codimension two bifurcation.

\section{A Reduced Representation}
\label{sec:center_manifold}

One may evaluate not just the standard center manifold emanating from the fixed point of the new system, but also a parameter dependent (asymptotic) center manifold so that variations with Reynolds number can be obtained on a reduced representation of the system. Typically, one takes the inverse Reynolds number variation as a small parameter, $\inv{\Rey} = \invRec(1 - \epsilon)$, or some variation thereof, accounting for scaling \citep{sipp07,carini15,negi24}. In this case we would also like to consider the variation with respect to the newly introduced parameter $\sigma$. We would therefore consider a perturbation of this parameter as $\sigma = \sigma_{0} + \sigma'$, where $\sigma_{0} = -\mathfrak{R}(\lambda_{3,4})$.

Parametric (and asymptotic) center manifold evaluation can be performed in multiple different ways depending on one's preference and style. One could consider power series expansions in the amplitudes of the center subspace modes and then augment the power series with additional polynomial terms to account for the additional parameter perturbations (here $\epsilon$  and $\sigma'$), as has been done in \cite{coullet83,carini15}. Alternately, one could consider extended systems, and include additional trivial equations for the parameter evolutions, thereby promoting parameter perturbations to intrinsic center subspace modes of the extended system. This is an often used trick for small dynamical systems, with some applications to larger systems as well \citep{mercer90,cox91,negi24,vizzaccaro24}. Recently, asymptotic expansions of spectral submanifolds of extended systems was performed in \cite{vizzaccaro24} in the context of external forcing. The consequences of system extension have been investigated to significant depth in the context of center manifolds in \cite{negi26}. The extended systems approach was taken here. The methodology is described here only conceptually. One may find detailed exposition of asymptotic evaluation of invariant manifolds in several works in the literature, see for example,  \cite{coullet83,roberts97,wiggins03,carini15,jain22,vizzaccaro24,negi24,negi26}.  

Briefly, the solution of equation~\eqref{eqn:compact_evolution2} restricted to the center manifold is assumed to be, $\vecubar(t) = \mathcal{Y}(\vecparv(t))$, where, $\vecparv$ is the vector of dimensionality equal to that of the center subspace (of the extended system). The time evolution of $\vecparv$ is assumed to be, $\dt{\vecparv} = \mathcal{G}(\vecparv)$, where, $\mathcal{G}$ is the sought after reduced representation of the system dynamics. Both $\mathcal{Y}$ and $\mathcal{G}$ are then expanded as a power series in $\vecparv$,
\begin{subequations}
\begin{align}
	\mathcal{Y}(\vecparv) = \sum \vecybar_{i}\parv_{i} + \sum \sum\vecybar_{i,j}\parv_{i}\parv_{j} \ldots, \\ 
	\mathcal{G}(\vecparv) = \sum \mathbf{g}_{i}\parv_{i} + \sum \sum\mathbf{g}_{i,j}\parv_{i}\parv_{j} \ldots,
\end{align}
\end{subequations}
and their substitution into equation~\eqref{eqn:compact_evolution2} results in a series of homological equations, that can be solved order by order. To obtain the reduced representation in its normal form, all entries in the asymptotic expansion of $\mathcal{G}$ are set to zero, except those that are required to remove singularities due to resonance in the homological equation. 

This results in the normal form of the parametric center manifold obtained for the synthetic system as,
\begin{subequations}
	\label{eqn:normal_form}
	\begin{align}
		\label{eqn:normal_form1}
			\dt{\parv_{1}} =& \left(g^{1}_{1} 
			+ g^{1}_{1,3,4}|\parv_{3}|^{2} + 	g^{1}_{1,1,2}|\parv_{1}|^{2}\right)\parv_{1}, \\
		\label{eqn:normal_form3}
			\dt{\parv_{3}} =& \left(g^{3}_{3}  
			+ g^{3}_{3,3,4}|\parv_{3}|^{2}
			+ g^{3}_{1,2,3}|\parv_{1}|^{2}\right)\parv_{3},
	\end{align}
\end{subequations}
where,
\begin{equation}
\begin{aligned}
	g^{1}_{1} =& 7.495\iota + (0.835 + 0.724\iota)\epsilon 
	+ (0.325 + 0.230\iota)\epsilon^{2} \nonumber \\
	& - (0.004 - 0.001\iota)\epsilon\sigma', \nonumber \\
	g^{3}_{3} =& (\sigma' + 10.31\iota) + (1.801 + 1.096\iota)\epsilon \nonumber \\
	& + (0.631 + 0.395\iota)\epsilon^{2} - (0.005 - 0.001\iota)\epsilon\sigma', \nonumber 
\end{aligned}
\end{equation}
\begin{equation}
	\begin{aligned}
		g^{1}_{1,1,2} = -573 + 340\iota, && g^{1}_{1,3,4} = -1553 - 342\iota, \nonumber \\ 
		g^{3}_{1,2,3} = - 829 + 275\iota, && g^{3}_{3,3,4} =  -751 - 6.99\iota. \nonumber
	\end{aligned}
\end{equation}
Here, $\parv_{1}$ refers to the amplitude corresponding to the $\widetilde{\lambda}_{1}$ mode and $\parv_{3}$ refers to the amplitude of the $\widetilde{\lambda}_{3}$ mode. The equations for $\parv_{2}$ and $\parv_{4}$ are the corresponding complex conjugates of equation~\ref{eqn:normal_form}. As mentioned earlier, $\epsilon$ is the perturbation parameter for the Reynolds number, defined using $\inv{\Rey} = \invRec(1 - \epsilon)$. If we consider $\sigma' = - \sigma_{0}$, the additional terms that were introduced to generate the synthetic system \eqref{eqn:compact_evolution2} vanish, 
and one obtains the original system \eqref{eqn:compact_evolution1} that had a codimension one bifurcation point. The first term of $g^{3}_{3}$ in equation~\eqref{eqn:normal_form3} becomes $(-0.092 + 10.31\iota)$, which is precisely the $\lambda_{3}$ eigenvalue of the original system. Henceforth the value of $\sigma'=-\sigma_{0}$ is held fixed and it is treated simply as another constant coefficient term and the primary focus is on variations of $\epsilon$. 

\begin{figure*}
	\centering
	\includegraphics[width=0.32\textwidth]{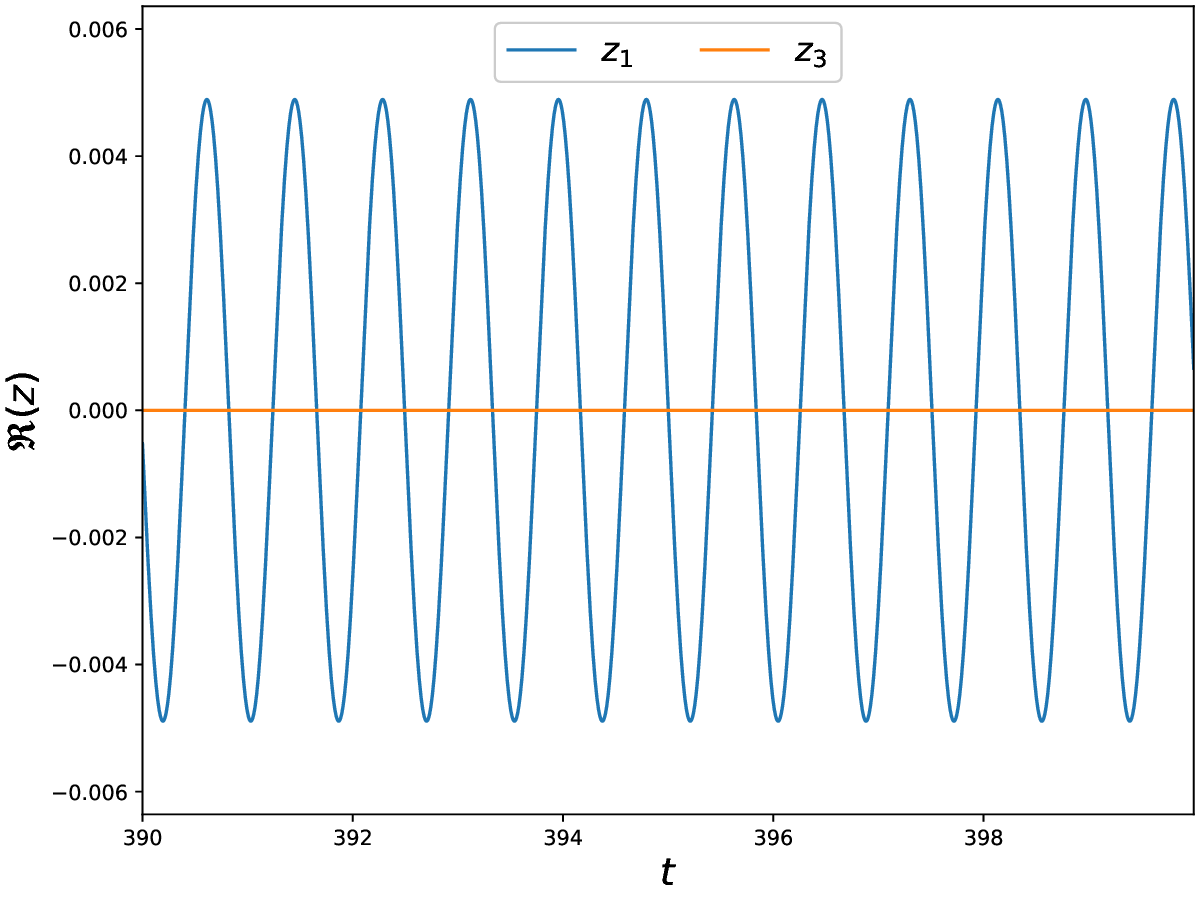}
	\includegraphics[width=0.32\textwidth]{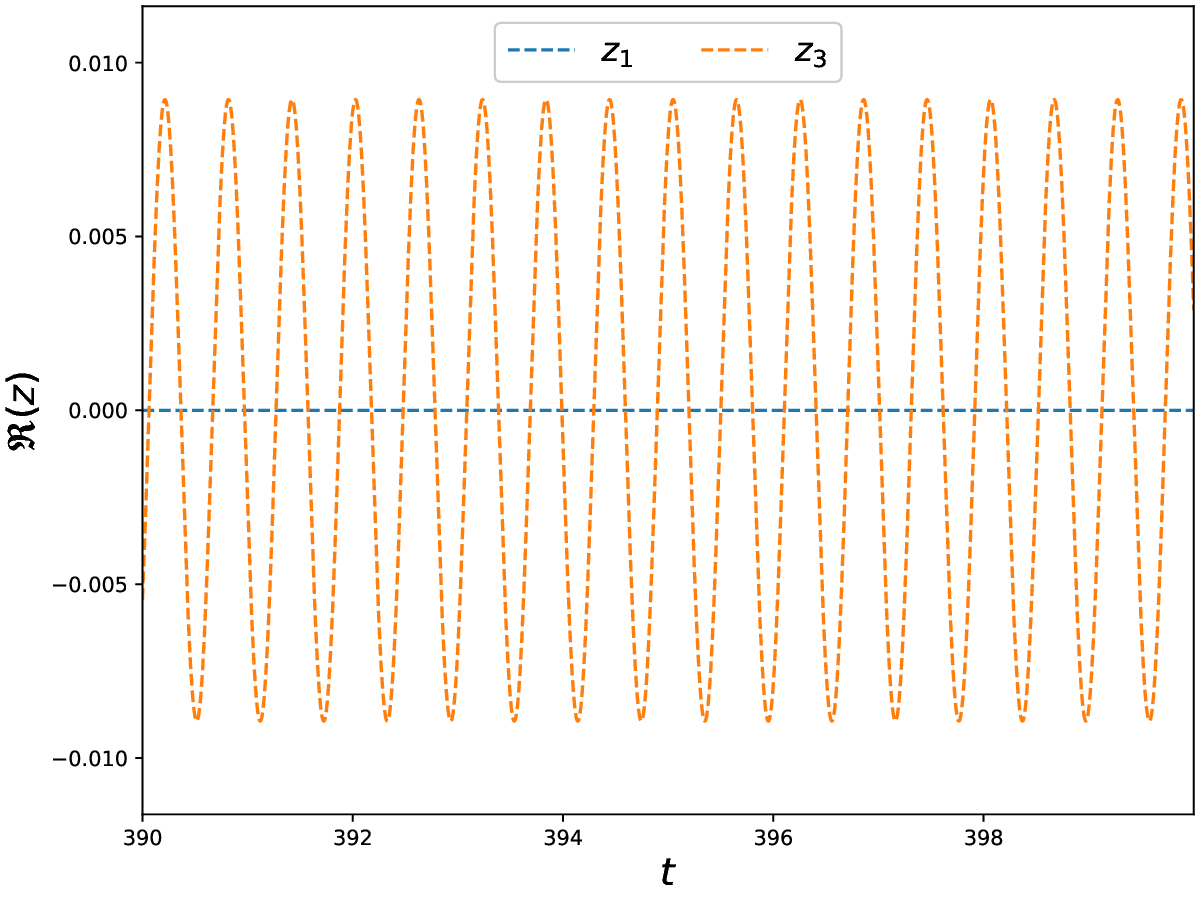}
	\includegraphics[width=0.32\textwidth]{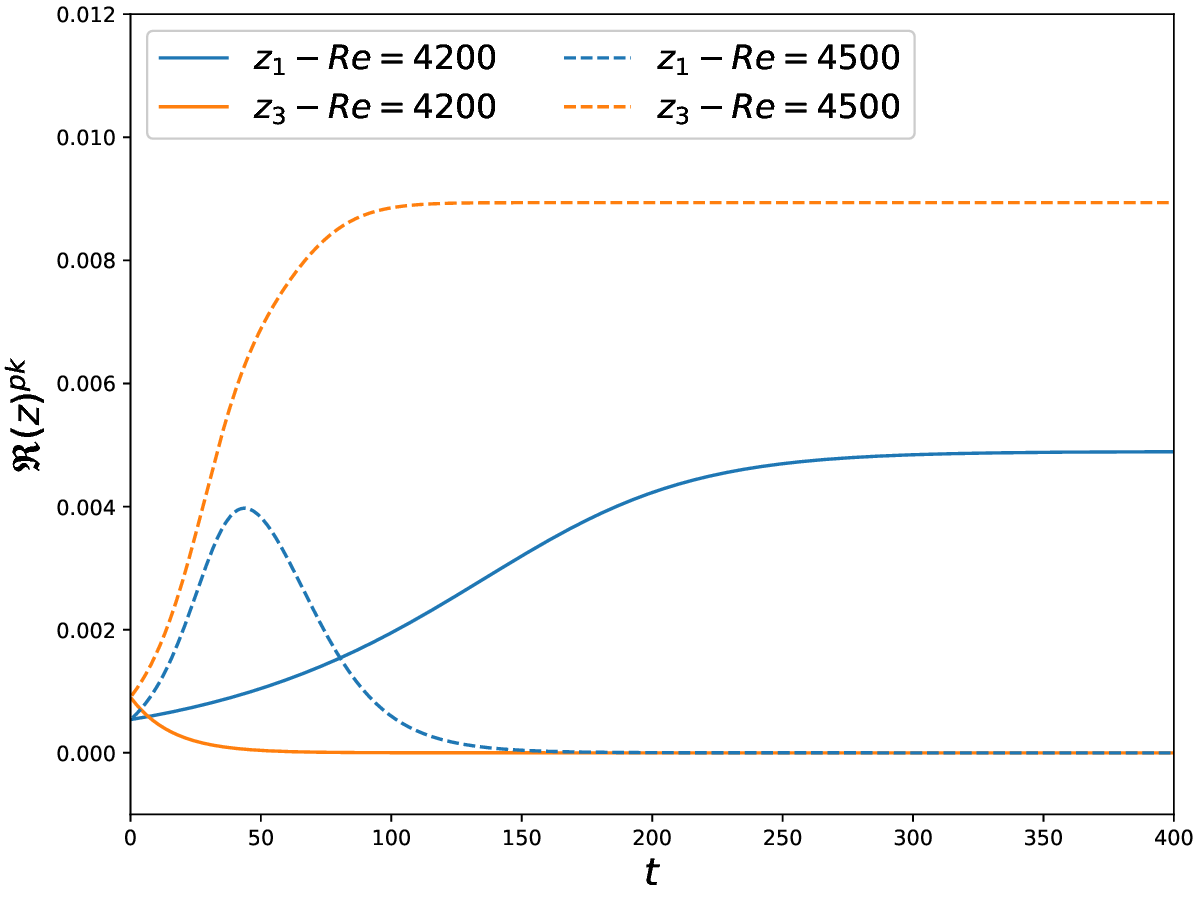}
	\caption{Real part of the time varying response of the reduced system at (left) $\Rey=4200$ and (center) $\Rey=4500$. The labels $\parv_{1}$ (blue), and $\parv_{3}$ (orange) correspond to the amplitudes of the modes $\lambda_{1}$ and $\lambda_{3}$ respectively. The figure on the right shows the time evolution of peak of the oscillation amplitudes for the two modes $\parv_{1},\parv_{3}$, and for the two different Reynolds numbers, $Re=4200$ (solid lines) and $Re=4500$ (dashed lines).}
	\label{fig:response}
\end{figure*}

The reduced system can be integrated in time to obtain the system response for different Reynolds numbers. This is plotted in figure~\ref{fig:response} for $\Rey=4200$, where one could expect the limit-cycle associated with $\lambda_{1,2}$ mode to emerge, and for $\Rey=4500$, where $\lambda_{3,4}$ mode limit-cycle could possibly emerge. The mode amplitudes $\parv_{1},\parv_{3}$ were given a small random initialization and the evolution is tracked. After a sufficiently long time the mode amplitude $\parv_{3}$ has decayed to zero and $\parv_{1}$ dominates the system response for $\Rey=4200$. On the other hand, for $\Rey=4500$, the situation is reversed and $\parv_{3}$ dominates the system response and $\parv_{1}$ has decayed to zero. The final system frequencies are then determined by the dominating modes at large times. The peak value of the individual mode oscillations is plotted for the entire time history in the right most panel in figure~\ref{fig:response}. The opposing evolution of the two modes at the two different Reynolds numbers is clearly visible.

\begin{figure*}
	\centering
	\includegraphics[width=.98\textwidth]{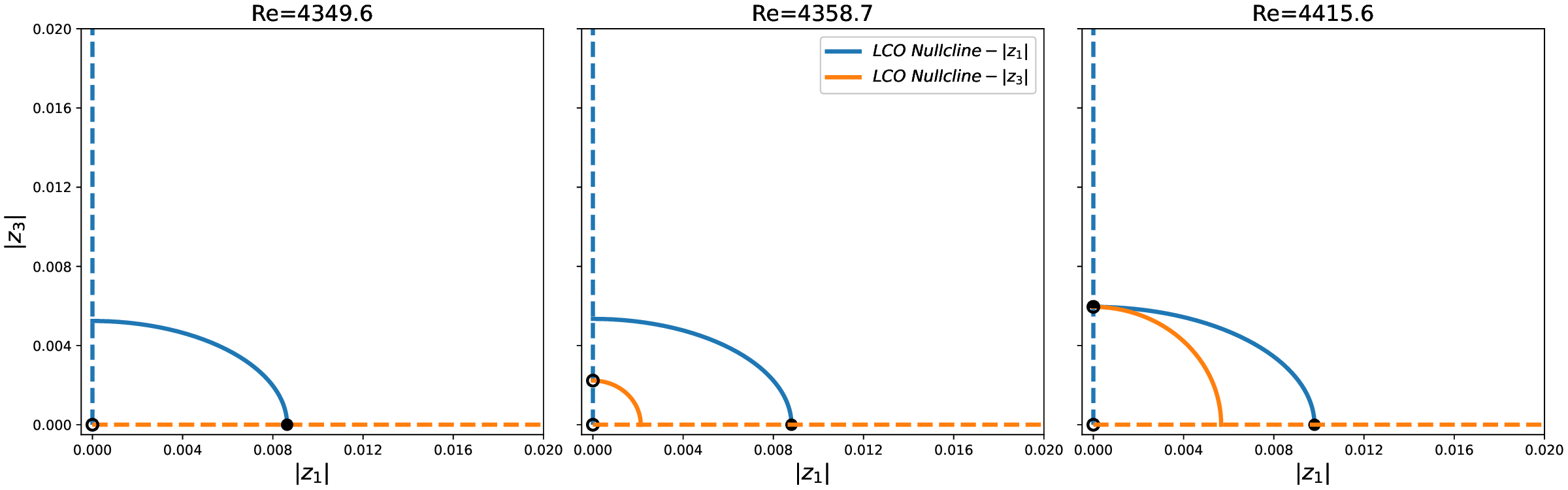}
	\includegraphics[width=.98\textwidth]{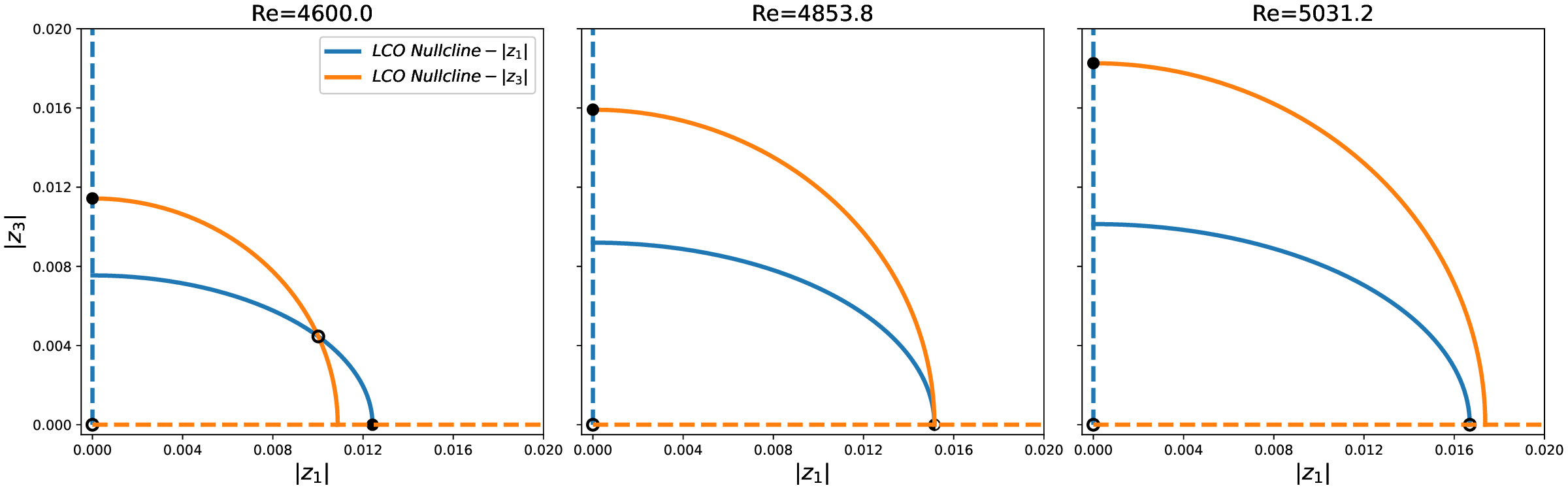}
	\caption{Evolution of the LCO null-clines as the Reynolds number is increased past the first bifurcation point. The solid blue line indicates the null-cline of the $\lambda_{1,2}$-LCO emerging from the first bifurcation. The solid orange line indicates the null-clines of the $\lambda_{3,4}$-LCO emerging from the second bifurcation point. The intersection of these LCO null-clines produces the quasi-periodic edge-state. }
	\label{fig:bifurcations}
\end{figure*}

To further determine the points where the mode switching takes place, one can obtain the evolution equations for the squared amplitudes, $|\parv_{1}|^{2}$ and $|\parv_{3}|^{2}$ from equation~\eqref{eqn:normal_form}, by multiplying each equation by its conjugate variable, \ie, we multiply equation~\eqref{eqn:normal_form1} by $\parv^{*}_{1}$ and equation~\eqref{eqn:normal_form3} by $\parv^{*}_{3}$. Utilizing the conjugate equations allows us to get rid of the imaginary terms and one obtains the equations for the squared amplitudes as,
\begin{subequations}
	\label{eqn:amplitude_equations}
	\begin{equation}
		\label{eqn:amplitude1}
		\begin{aligned}
			\dfrac{d |\parv_{1}|^{2}}{dt} =& 2\left(\mathfrak{R}(g^{1}_{1}) + \mathfrak{R}(g^{1}_{1,3,4})|\parv_{3}|^{2} + \mathfrak{R}(g^{1}_{1,1,2})|\parv_{1}|^{2}\right)|\parv_{1}|^{2}
		\end{aligned}
	\end{equation}
	\begin{equation}
		\label{eqn:amplitude3}
		\begin{aligned}
			\dfrac{d |\parv_{3}|^{2}}{dt} =& 2\left(\mathfrak{R}(g^{3}_{3}) + \mathfrak{R}(g^{3}_{3,3,4})|\parv_{3}|^{2} + \mathfrak{R}(g^{3}_{1,2,3})|\parv_{1}|^{2}\right)|\parv_{3}|^{2},
		\end{aligned}
	\end{equation}
\end{subequations}
where, $\mathfrak{R}(\cdot)$ represents the real part of a quantitiy. These equations can be analyzed for equilibrium when the time derivatives vanish. Since both $\epsilon$ and $\sigma'$ are just treated as constant parameters, equation~\eqref{eqn:amplitude_equations} can be analyzed by parametrically plotting the null-clines of the evolution functions on the right hand sides of the two equations, while restricting the analysis to the first-quadrant of the $|\parv_{1}|-|\parv_{3}|$ phase plane. Obviously, $|\parv_{1}| = 0$ and $|\parv_{3}| = 0$ are the trivial null-clines for equations~\eqref{eqn:amplitude1} and \eqref{eqn:amplitude3} respectively. These represent the complete absence of the respective modes at equilibrium, and are marked using dashed lines in figure~\ref{fig:bifurcations}. The non-trivial null-clines, representing LCO amplitudes are given by,
\begin{subequations}
	\label{eqn:nullclines}
	\begin{equation}
		\label{eqn:nullcline1}
		\begin{aligned}
			 \mathfrak{R}(g^{1}_{1}) + \mathfrak{R}(g^{1}_{1,3,4})|\parv_{3}|^{2} + \mathfrak{R}(g^{1}_{1,1,2})|\parv_{1}|^{2} = 0,
		\end{aligned}
	\end{equation}
	\begin{equation}
		\label{eqn:nullcline3}
		\begin{aligned}
			\mathfrak{R}(g^{3}_{3}) + \mathfrak{R}(g^{3}_{3,3,4})|\parv_{3}|^{2} + \mathfrak{R}(g^{3}_{1,2,3})|\parv_{1}|^{2} = 0.
		\end{aligned}
	\end{equation}
\end{subequations}
These non-trivial null-clines will be referred to as the LCO null-clines and are represented using solid lines in figure~\ref{fig:bifurcations}. At $\epsilon=0$, representing the first bifurcation point, there are no solutions for equation~\eqref{eqn:nullcline3}. Therefore at bifurcation, no limit-cycle for the $\lambda_{3,4}$ mode exists. For equation~\eqref{eqn:nullcline1}, $\mathfrak{R}(g^{1}_{1})=0$, and a trivial solution $\parv_{1}=0$, $\parv_{3}=0$ exists. This of course represents the birth of the first limit-cycle associated with the $\lambda_{1,2}$ mode, as one would expect. \cite{bengana19} refer to this as $\Rey_{2}$. The authors have numbered the important bifurcation events based on the closest limit cycle that can be associated with the event. Here, we follow the numbering sequentially in order of its occurrence as $\epsilon$ is increased. To avoid confusion, we will refer to the bifurcation events in this study with an overhead tilde, so the first bifurcation point will be refered to as $\widetilde{\Rey}_{1}$. As $\epsilon$ is increased a non-trivial solution of $|\parv_{1}|$ can be found, with $|\parv_{3}|=0$ (trivial nullcline), which represents the growing $\lambda_{1,2}$-LCO as the system moves away from the bifurcation point.

The next interesting point occurs when $\lambda_{3,4}$-LCO is born, which occurs when the solution for equation~\eqref{eqn:nullcline3} exists for the first time. This occurs at, 
$\widetilde{\Rey}_{2} = 4349.6$. \cite{bengana19} refer to this as $\Rey_{3}$ and obtain a value of $4348$ through a quadratic approximation, which is remarkably close to the value obtained here. The null-clines for the system are shown in the top left panel in figure~\ref{fig:bifurcations}. The intersection of the null-clines are referred to as \emph{equilibrium points}, since they represent simultaneous equilibrium of the limit cycles. These are marked using circles, with stable equilibrium points denoted by filled circles, while unstable ones denoted by empty circles. Further increases in Reynolds number result in two distinct LCO null-clines however, these LCO null-clines intersect only with the trivial null-clines, $|\parv_{1}|=0$ or $|\parv_{3}|=0$. Therefore no invariant quasi-periodic (QP) solutions exist so far. The bifurcation to a QP solution occurs at $\widetilde{\Rey}_{3}=4415.6$, when the two LCO null-clines first intersect (top right panel in figure~\ref{fig:bifurcations}). This corresponds to $\Rey'_{3}=4410$ in \cite{bengana19}, which is again close to the value obtained here. In \cite{bengana19} the authors find that around this bifurcation the characteristics of the quasi-periodic state are much closer to those of the limit-cycle of the second bifurcation, which the authors refer to as $LC_{3}$ ($\lambda_{3,4}$-LCO in the current study). This is indeed what is found here as well. The first intersection of the two LCO null-clines occurs for a vanishing value of $|\parv_{1}|$. Therefore, at the inception of the QP state the $\lambda_{1,2}$-LCO has a vanishingly small amplitude and the QP state characteristics will be dominated by those of the $\lambda_{3,4}$-LCO.

The QP solutions have been identified as the edge state between the two limit cycles in \cite{bengana19}. As the Reynolds number is further increased, the edge-state has non-trivial components of both the $\parv_{1}$ and $\parv_{3}$ mode amplitudes. The edge-state moves along the intersection of the null-clines until, at $\widetilde{\Rey}_{4}=4853.8$, the edge state is such that it has a vanishing $\parv_{3}$ component (bottom middle panel). The QP state ceases to exist beyond this point. In \cite{bengana19} the disappearance of the of the edge state is found to be at around $\Rey'_{2}=4600$, which is a bit different from what is predicted in the current study. Beyond $\widetilde{\Rey}_{4}$, there exist two distinct equilibrium points of the system, one for each limit cycle, although the $\lambda_{1,2}$-LCO is unstable. 

The role of the cubic terms in equation~\eqref{eqn:normal_form}, or the quartic terms in equation~\eqref{eqn:amplitude_equations} is to provide a saturation mechanism for a growing limit-cycle (when the coefficient of these terms are negative). For a standard Hopf bifurcation, only a self saturation term exists, $|\parv_{1}|^{4}$ for equation~\eqref{eqn:amplitude1} for example, and the cross interaction term $|\parv_{1}|^{2}|\parv_{3}|^{2}$ is absent. The cross-interaction terms in both equations~\eqref{eqn:amplitude1} and ~\eqref{eqn:amplitude3} have a negative sign so that the presence of one mode strengthens the saturation mechanism of the other. Earlier, we identified the second bifurcation in the flow as the first non-trivial existence of the $\lambda_{3,4}$-LCO, which is the top left panel in figure~\ref{fig:bifurcations}. However, this bifurcation occurs when $|\parv_{1}|=0$. If we imagine a commonly employed strategy for analyzing systems, which is to slowly vary a system parameter and observe the response, the bifurcation scenario looks different. Increasing the Reynolds number marginaly from the first bifurcation point and allowing the system to equilibriate, means that the system reaches the equilibrium point lying of the $|\parv_{1}|$ axis in figure~\ref{fig:bifurcations}. The system then follows the equilibrium point with slow increases of the Reynolds number. At the second bifurcation point, $\widetilde{\Rey}_{2}$, the $\lambda_{3,4}$-LCO would emerge (at least transiently) for a quiescent flow state, however, the non zero $|\parv_{1}|$ leads to added damping effects and the $\lambda_{3,4}$-LCO remains non-existant. In this scenario, the $\lambda_{3,4}$ LCO can only emerge when the \emph{effective} eigenvalue for the $\parv_{3}$ evolution starts having a non-negative real part. This cross-over point can be found by looking at the Reynolds number dependent linear multiplier of $|\parv_{3}|^{2}$ in equation~\eqref{eqn:amplitude3}, \ie, 
\begin{align}
	d|\parv_{3}|^2/dt = \alpha_{3}(\epsilon)|\parv_{3}|^{2}, \nonumber \\
	\alpha_{3}(\epsilon) = 2\left(\mathfrak{R}(g^{3}_{3}) +  \mathfrak{R}(g^{3}_{1,2,3})|\parv_{1}(\epsilon)|^{2}\right), \nonumber
\end{align}
where, $|\parv_{1}(\epsilon)|$, is the Reynolds number dependent equilibrium value of $|\parv_{1}|$, for $|\parv_{3}|=0$. When $\alpha_{3}(\epsilon)$ first becomes non-negative, is the new ``path dependent'' bifurcation point. This is found to coincide with $\widetilde{\Rey}_{4}=4853.8$. This was found using the squared amplitude equations~\eqref{eqn:amplitude_equations} however, using equation~\eqref{eqn:normal_form} and calculating the effective eigenvalue of $\parv_{3}$ by considering $|\parv_{1}(\epsilon)|^{2}$ constant leads to the same bifurcation value. At this point, the $\lambda_{1,2}$-LCO loses stability since, any increase in $\parv_{3}$ strengthens the saturation mechanism provided by the cross interaction terms for $\lambda_{1,2}$-LCO, and therefore reduces the equilibrium value of $|\parv_{1}|$. This reduction in $|\parv_{1}|$ implies the effective growth rate for $\parv_{3}$ becomes larger and $|\parv_{3}|$ further increases. Thus a feedback loop is established, moving the system away from the equilibrium point on the $|\parv_{1}|$ axis, ending up at equilibrium point on the $|\parv_{3}|$ axis, and the system now exhibits the $\lambda_{3,4}$-LCO. The system now follows this new equilibrium point with further slow increases in Reynolds number. 

The phenomenon repeats itself, with the roles of the two limit cycles reversed, when the system is slowly varied in the opposite direction. For a high enough Reynolds number the $\lambda_{3,4}$-LCO exists and the system is at the equilibrium point on $|\parv_{3}|$ axis. Again, the non-zero $|\parv_{3}|$ causes $\parv_{1}$ to have a damped effective eigenvalue, thereby suppressing the emergence of the $\lambda_{1,2}$-LCO. The effective growth rate along the decreasing Reynolds number path is again predicted using the terms linear in $|\parv_{1}|^{2}$ in equation~\eqref{eqn:amplitude1}, 
\begin{align}
	\alpha_{1}(\epsilon) =& 2\left(\mathfrak{R}(g^{1}_{1}) + \mathfrak{R}(g^{1}_{1,3,4})|\parv_{3}(\epsilon)|^{2}\right), \nonumber
\end{align}
where $|\parv_{3}(\epsilon)|$ is the Reynolds number dependent equilibrium value of $|\parv_{3}|$. As one might anticipate, $\alpha_{1}(\epsilon)$ becomes non-negative at $\widetilde{\Rey}_{3}$ (top right panel in figure~\ref{fig:bifurcations}) and the same feedback loop described earlier, now causes the system to move away from the $\lambda_{3,4}$-LCO and reach the $\lambda_{1,2}$-LCO equilibrium point. The $\lambda_{1,2}$-LCO remains stable below $\widetilde{\Rey}_{3}$ until the first bifurcation point is reached. Between $\widetilde{\Rey}_{3}$ and $\widetilde{\Rey}_{4}$ the system exhibits bistability and a classic hysterisis phenomenon is obtained. The mechanism for the exchange of stability between the two limit cycles is establised through the cross-interaction term, and the change in sign of the effective growth rate of one of the modes when the system is at equilibrium of the other modes' limit cycle. 

Usually, this phenomenon is predicted by undertaking a Floquet analysis of the existing limit cycle, as has been done in \cite{bengana19}. Floquet analysis on the reduced system \eqref{eqn:normal_form} confirms that the two limit-cycles indeed become unstable at $\widetilde{\Rey}_{3}$ and $\widetilde{\Rey}_{4}$. The LCO null-clines in figure~\ref{fig:bifurcations} provides the geometrical structure underlying the exchange of stabilities while the reasoning with the effective eigenvalues and feedback loop due to the cross-interaction saturation terms gives the physical explanation of the process.

\begin{figure*}
	\centering
	\includegraphics[width=.49\textwidth]{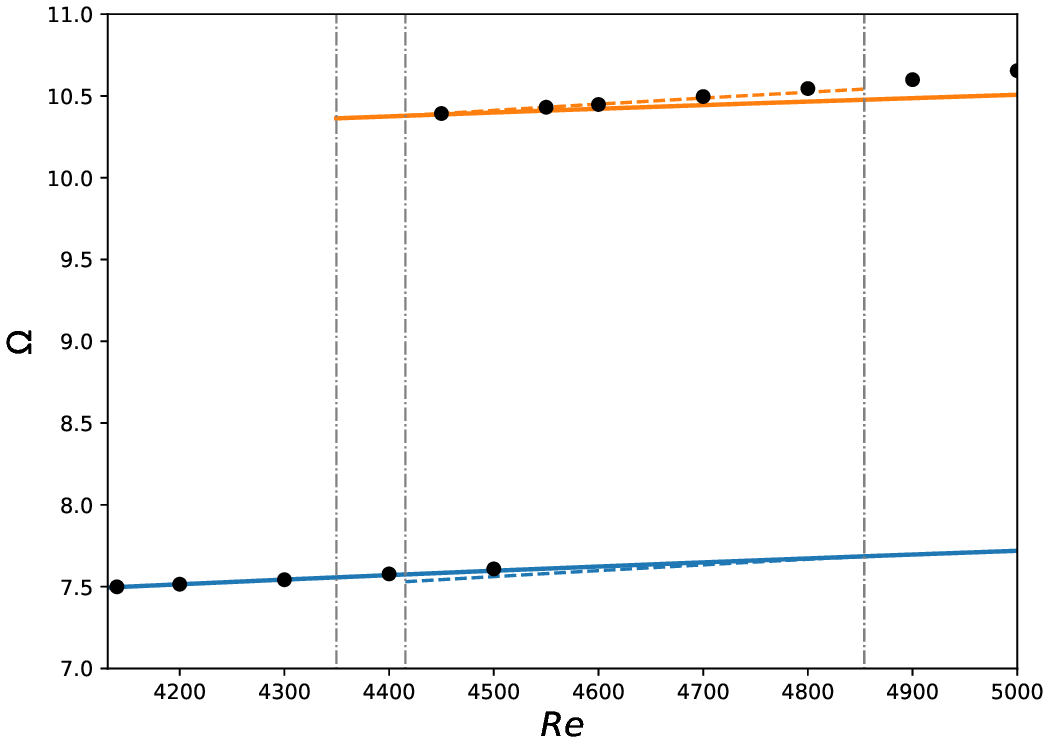}
	\includegraphics[width=.49\textwidth]{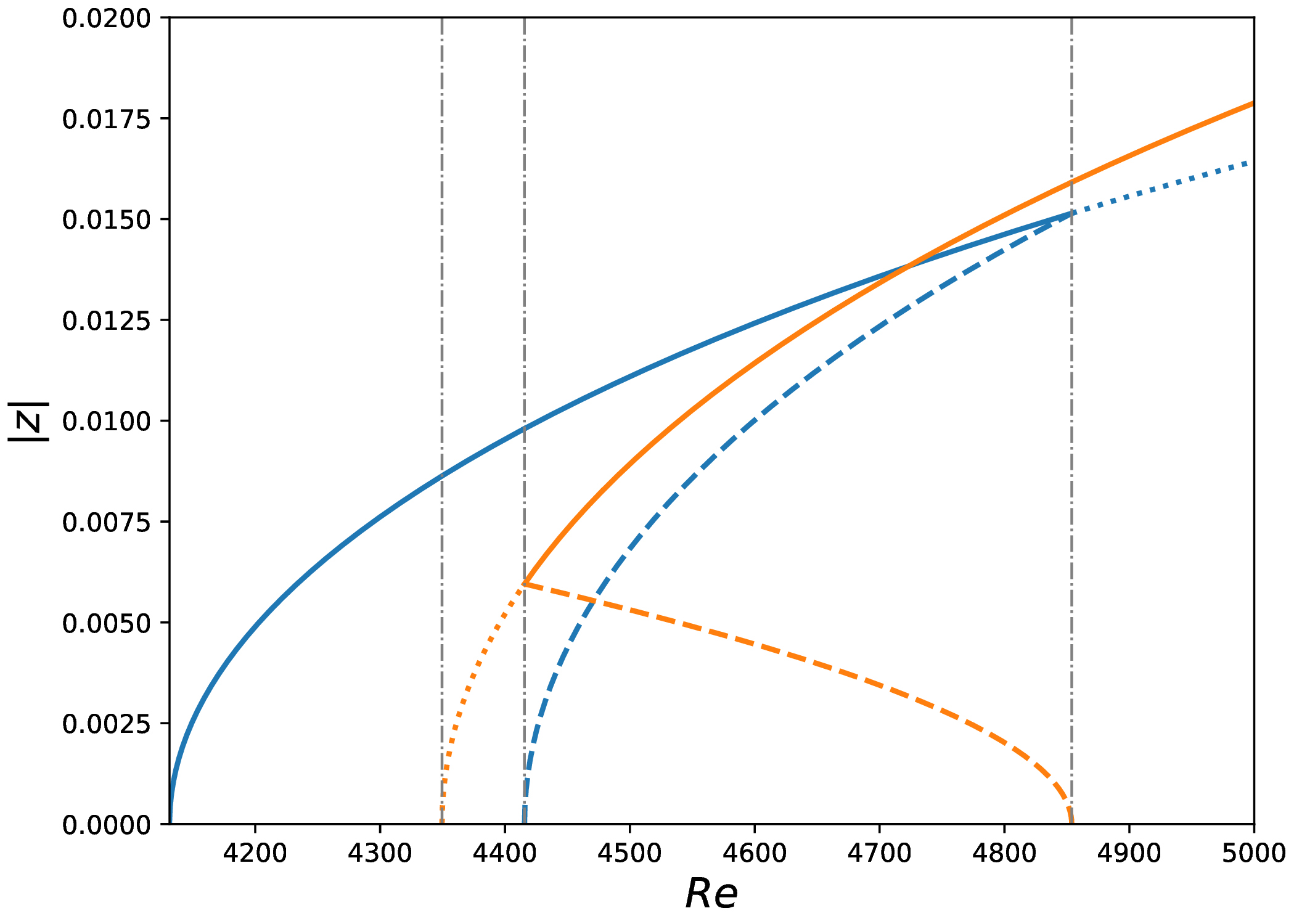}
	\caption{(Left) Comparison of the angular frequencies of the full system and the reduced model. The blue line indicates the $\lambda_{1,2}$ LCO angular frequencies while the orange line indicates the $\lambda_{3,4}$ LCO angular frequencies. The solid black circles indicates the frequencies obtained from non-linear simulations. And (right), the bifurcation diagram obtained in terms of the equilibrium LCO amplitudes $|\parv|$. The blue lines indicate $|\parv_{1}|$ while the orange line indicates $|\parv_{3}|$. The solid lines indicate stable equilibria amplitudes while the dotted part of the lines indicate unstable equilibria amplitudes. The dashed lines indicate amplitudes for the (unstable) QP state. The $-\cdot$ gray vertical lines mark the successive bifurcations points identified in the work ($\widetilde{\Rey}_{2}=4349.6$, $\widetilde{\Rey}_{3}=4415.6$ and $\widetilde{\Rey}_{4}=4853.8$).}
	\label{fig:equilibrium_omega}
\end{figure*}

Finally, since the equilibrium points are known through the reduced equations, the equilibrium frequencies can also be predicted. This is shown in figure~\ref{fig:equilibrium_omega} (left). The blue line indicates the angular frequencies for the $\lambda_{1,2}$-LCO equilibrium point, while the orange line indicates the angular frequencies for the $\lambda_{3,4}$-LCO. The dashed lines indicate the angular frequencies for the respective modes at the quasi-periodic edge state. The filled black circles indicate the results obtained through the non-linear simulations. The $\lambda_{1,2}$-LCO angular frequencies are well predicted while a small deviation exists for the $\lambda_{3,4}$-LCO at higher Reynolds numbers. As indicated in the work of \cite{meliga17}, these are likely due to contributions of higher order terms which have not been accounted for in the current work.  The bifurcation diagram in terms of the mode amplitudes $|\parv_{1}|, |\parv_{3}|$ is also shown in figure~\ref{fig:equilibrium_omega} (right) as a function of Reynolds number, indicating different regions of LCO stability (solid lines), instability (dotted lines) and QP states (dashed lines) as predicted by the model. 

\section{Conclusion}
\label{sec:conclusion}

A reduced mathematical model for the flow in an open cavity is derived through a center manifold reduction of a perturbed system. The original system is then approximated asymptotically, along with the Reynolds number and center subspace mode variations, resulting in a normal form of a (Reynolds number dependent) double Hopf bifurcation. The model exhibits many of the key characteristics of the system found in the detailed study by \cite{bengana19}, including existence of bistability, switching of limit cycles and a quasi-periodic edge state between the two limit cycles. The reduced model provides a mechanism for the exchange of the stability of the two limit-cycle oscillations based on the cross-interaction term and effective eigenvalue of the modes at the equilibrium points of the limit-cycles. The reduced model provides a good prediction of the two secondary bifurcation points $\widetilde{\Rey}_{2}$ and $\widetilde{\Rey}_{3}$, corresponding to the first inception of the $\lambda_{3,4}$-LCO and QP states. However, $\widetilde{\Rey}_{4}$ is overestimated as compared to the results of \cite{bengana19}. The author suspects this may be due to the asymptotic nature of the approximation, which typically deviates systematically as the system is moved away from the (first) bifurcation point. Nevertheless the overall qualitative structure of the solutions appears to be well captured by the model. 

\section*{Acknowledgements}
The author is grateful for the computational resources and support provided by the Scientific Computing and Data Analysis section of Research Support Division at OIST. The author would also like to thank Dr. Ardeshir Hanifi for his comments on the manuscript.

The research was supported by the Okinawa Institute of Science and Technology Graduate	University (OIST) with subsidy funding from the Cabinet Office, Government of Japan.

\FloatBarrier

\bibliographystyle{apalike}

\end{document}